\documentclass[aps,prc,preprint,amsmath,amssymb,showpacs,preprintnumbers,superscriptaddress]{revtex4-1}
\usepackage{CJK}
\usepackage{graphicx}% Include figure files
\usepackage{dcolumn}% Align table columns on decimal point
\usepackage{bm}% bold math
\usepackage{color}
\usepackage{hyperref}% add hypertext capabilities
\allowdisplaybreaks

\begin{document}

\title{De-excitation effects on entanglement in multi-nucleon transfer reactions}% 

\author{Y. C. Yang}
\affiliation{State Key Laboratory of Nuclear Physics and Technology, School of Physics, Peking University, Beijing 100871, China}

\author{D. D. Zhang}
\affiliation{Institute of Theoretical Physics, Chinese Academy of Science, Beijing 100871, China}

\author{D. Vretenar}
\email{vretenar@phy.hr}
\affiliation{Physics Department, Faculty of Science, University of Zagreb, 10000 Zagreb, Croatia}
\affiliation{State Key Laboratory of Nuclear Physics and Technology, School of Physics, Peking University, Beijing 100871, China}

\author{B. Li}
\affiliation{State Key Laboratory of Nuclear Physics and Technology, School of Physics, Peking University, Beijing 100871, China}

\author{T. Nik\v{s}i\'{c}}
\affiliation{Physics Department, Faculty of Science, University of Zagreb, 10000 Zagreb, Croatia}
\affiliation{State Key Laboratory of Nuclear Physics and Technology, School of Physics, Peking University, Beijing 100871, China}

\author{P. W. Zhao}
\email{pwzhao@pku.edu.cn}
\affiliation{State Key Laboratory of Nuclear Physics and Technology, School of Physics, Peking University, Beijing 100871, China}

\author{J. Meng}
\email{mengj@pku.edu.cn}
\affiliation{State Key Laboratory of Nuclear Physics and Technology, School of Physics, Peking University, Beijing 100871, China}

\date{\today}

\begin{abstract}
This study quantifies the impact of nuclear de-excitation on correlations in multi-nucleon transfer (MNT) reactions. To bridge the gap between initial collision dynamics and final experimental observables, we introduce a hybrid TDCDFT+GEMINI approach, integrating time-dependent covariant density functional theory (TDCDFT) with the statistical de-excitation model GEMINI++. Applied to the $^{40}$Ca + $^{208}$Pb reaction, our method demonstrates that the de-excitation is essential for reconciling theoretical cross sections with experimental data. Analysis of the cross-section Shannon entropy reveals that new reaction channels open abruptly at a specific energy threshold. By employing mutual information, we show that the de-excitation process significantly degrades the initial quantum entanglement between the projectile-like and the target-like fragments, revealing a key mechanism through which fundamental quantum correlations are lost.
\end{abstract}

\maketitle

%\tableofcontents

\newpage
%--------------------------------------------------------------

\section{Introduction} \label{intro}
%--------------------------------------------------------------

Multi-nucleon transfer (MNT) reactions are a class of nuclear processes in which two heavy nuclei collide in a slow, ``grazing" encounter, resulting in the exchange of multiple nucleons (protons and neutrons). These reactions have emerged as a unique tool in nuclear physics, vital for producing neutron-rich nuclei~\cite{Zagrebaev2008PRL, Watanable2015PRL, Niwase2023PRL}, synthesizing heavy and superheavy elements~\cite{Zagrebaev2006PRC, Zagrebaev2007JPGNPP, Zagrebaev2008PRC, Zagrebaev2011PRC}, and probing nuclear structure and reaction mechanisms. Consequently, MNT reactions have been listed as one of the primary research topics at many nuclear reaction facilities~\cite{Rotaru2022NIMPRB, Huang2017, GALES2007PPNP, Mijatovic2022FP, Corradi2013NIMPR}. Furthermore, a variety of theoretical models, including the GRAZING model~\cite{WintherNPA1994,Winther1995NPA,PWWEN2019PRC},  the dinuclear system (DNS)
model~\cite{Feng2017PRC,Niu2017PRC,Zhu2017PRC,XJBao2018PLB,Zhu2018PRC,Guo2019PRC,PHChen2022PRC,YHZhang2023PRC,ZQFeng2023PRC,ZHLiao2023PRC,ZHLiao2023PRR,TLZhao2023PRC,TLZhao2024PRC,GZhang2025PRC}, the Langevin equations~\cite{Zagrebaev2008PRL,Zagrebaev2008PRC,Zagrebaev2011PRC,Zagrebaev2013PRC,AVKarpov2017PRC,VVSaiko2019PRC}, the improved quantum molecular dynamics (ImQMD) model ~\cite{ZHLiao2023PRC,KZhao2015PRC,Cli2016PRC,HYao2017PRC,Cli2019PRC,KZhao2021PLB,KZhao2022PRC,CLi2024PRC,PFeng2025PRC}, the time-dependent density functional theory (TDDFT)~\cite{Simenel2010PRL,Sekizawa2013PRC,Sekizawa2014PRC,Sekizawa2016PRC,Sekizawa2017PRC,Sekizawa2017PRC(2),ZJWu2019PRC,XJiang2020PRC,ZJWu2022PLB,DDZhang2024PRC,DDZhang2024PRC2,DDZhang2025PLB}, and beyond-mean-field method~\cite{Ocal2025PRC, ZGao2025PRC}, have been developed in order to explore the optimal projectile-target combination, incident energy, and related conditions for the synthesis of neutron-rich and superheavy nuclei.

Experimental evidence shows that the de-excitation processes in MNT reactions, including particle evaporation and fission, significantly alter the yield distribution of heavy primary nuclei, shifting it toward lighter isotopes~\cite{Corradi1999PRC,mijatovi2016PRC,Watanable2015PRL,Vogt2015PRC,Szilner2007PRC,John2014PRC,John2017PRC,Kratz2013PRC,Kuzulin2012PRC,Kuzulin2016PRC,Galtarossa2018PRC}. Consequently, incorporating these mechanisms is essential for theoretical models to provide reliable predictions. Taking incident energy as an example, higher energies may enhance the production cross sections of heavy primary nuclei but also leave them in higher excitation energies. Therefore, whether these products survive the subsequent de-excitation process requires careful evaluation. Furthermore, the de-excitation processes weaken the correlation between the target-like fragments (TLFs) and the projectile-like fragments (PLFs), thereby complicating the identification of the TLFs. 

Experimentally, the PLFs can often be directly identified thanks to the large acceptance high resolution magnetic spectrometers~\cite{Stefanini2002NPA,Szilner2007PRC,MontanariEPJA2011,Rejmund2011NIMPRSA}. However, identifying the TLFs remains a significant challenge, despite various attempts to achieve this directly~\cite{Kratz2013PRC,Kuzulin2012PRC,Beliuskina2014EPJA,Kuzulin2016PRC} or indirectly~\cite{Corradi1999PRC,mijatovi2016PRC,Watanable2015PRL,Vogt2015PRC,Szilner2007PRC,John2014PRC,John2017PRC,Galtarossa2018PRC}. For instance, in the study of the $^{136}$Xe + $^{198}$Pt reaction at GANIL~\cite{Watanable2015PRL}, the cross sections of the TLFs were inferred by combining the identification of the light reaction partner and the detection of coincident $\gamma$-rays from the heavy partner. This approach constrains the mass and charge of the TLFs within a narrow range, though their exact values may still carry some uncertainty. This raises a key question: in MNT reactions, how much information about the TLFs can truly be deduced from measurements in coincidence with the PLFs?

In the previous study~\cite{BLi2024PRC2}, the entanglement between the primary PLFs and TLFs was investigated using time-dependent covariant density functional theory (TDCDFT). 
%The entanglement between primary PLFs and TLFs was measured via the mutual information of the von Neumann entropy, which provides an upper bound on the information contained in the correlation between any two observables pertaining to the PLFs and TLFs~\cite{Auletta2009}.
The entanglement was quantified through von Neumann entropy and its relationship
with nucleon-number fluctuation and particle number was analyzed.
Subsequently, the correlation of specific observable, namely spin, between the primary TLFs and PLFs was studied in Ref. ~\cite{DDZhang2025PLB}. The angular distribution of the fragments was calculated by incorporating angular moment projection and the correlations between the spins of thr primary fragments were quantified by mutual information. 

The analysis in Ref.~\cite{BLi2024PRC2,DDZhang2025PLB} was conducted for the primary fragments. How the correlation between the primary fragments is affected
by the de-excitation processes remains to be analyzed. In this work, we integrate the microscopic framework of TDCDFT with the statistical model GEMINI++~\cite{Charity2008} to study MNT reactions with full account of the de-excitation effects. The principal advantage of the TDCDFT framework~\cite{ZXRen2020PLB,ZXRen2020PRC,ZXRen2022PRC,ZXRen2022PRC2,ZXRen2022PRL,BLi2023PRC,BLi2024PLB,BLi2024PRC} is its ability to model the complete collision dynamics from first principles, without requiring ad hoc adjustments for specific reaction mechanisms. This combined approach is designated the TDCDFT+GEMINI method. The paper is structured as follows. Section II provides an outline of the TDCDFT+GEMINI methodology. Section III details the numerical aspects of the self-consistent TDCDFT calculations, particle number projection, and the GEMINI++ model. In Section IV, we present and discuss the results. Firstly, the MNT reaction $^{40}$Ca + $^{208}$Pb at $E_{\text {lab }}=249\ \mathrm{MeV}$ is analyzed as a benchmark; the results are compared with experimental data and previous TDHF+GEMINI calculations. We then investigate the energy dependence of the cross-section distributions by simulating reactions at incident energies from 235 to 270 MeV. Subsequently, we analyze the influence of the de-excitation on the correlations between the projectile-like and the target-like fragments, using mutual information as a quantitative measure. Finally, Section V offers a concise summary and concluding remarks.
%--------------------------------------------------------------
\newpage
\section{Theoretical framework } \label{Stheory}
%--------------------------------------------------------------

\subsection{Time-dependent covariant density-functional theory}
%--------------------------------------------------------------

In the adiabatic approximation of time-dependent density functional theory, and in the absence of pairing correlations, the nuclear wave function is at all times a Slater determinant of occupied single-particle states. The time evolution of the single-particle wave function $\psi_k(\boldsymbol{r}, t)$ is governed by the time-dependent Dirac equation~\cite{ZXRen2020PRC,Runge1984PRL,Leeuwen1999PRL}
\begin{equation}
i \hbar \frac{\partial}{\partial t} \psi_k(\boldsymbol{r}, t)=\hat{h}(\boldsymbol{r}, t) \psi_k(\boldsymbol{r}, t) .
\end{equation}
Here, the single-particle Hamiltonian $\hat{h}(\boldsymbol{r}, t)$ reads 
\begin{equation}
\hat{h}(\boldsymbol{r}, t)=\boldsymbol{\alpha} \cdot(\hat{\boldsymbol{p}}-\boldsymbol{V})+V^0+\beta(m+S),
\end{equation}
where $\boldsymbol{\alpha}, \beta$ are the Dirac matrices, and $m$ is the nucleon mass. The scalar potential $S$ and four-vector potential $\boldsymbol{V}$ are self-consistently determined at each step in time by the time-dependent densities and currents 
\begin{equation}
\begin{aligned}
\rho_S(\boldsymbol{r}, t) & =\sum_k \bar{\psi}_k \psi_k \\
j^\mu(\boldsymbol{r}, t) & =\sum_k \bar{\psi}_k \gamma^\mu \psi_k \\
j_{T V}^\mu(\boldsymbol{r}, t) & =\sum_k \bar{\psi}_k \gamma^\mu \tau_3 \psi_k
\end{aligned}
\end{equation}
For further details, we refer the reader to Refs.~\cite{ZXRen2020PRC,ZXRen2020PLB}.

Consider a nuclear collision between a projectile and a target, composed of $A^i_P,A^i_T$ nucleons, respectively. The system is initialized with a wave function formed from the direct product of the individual ground-state wave functions, boosted according to the relative velocity determined by the incident energy $E$ and impact parameter $b$. Following the collision, the system is assumed to separate into a projectile-like fragment (PLF) and a target-like fragment (TLF). The physical space can consequently be partitioned into two corresponding subspaces: $V^f_P$ for the PLF and $V^f_T$  for the TLF. The wave functions of these fragments are generally not eigenstates of the particle number operator but are instead superpositions of states with different nucleon numbers. To calculate the cross sections for specific reaction channels, we employ the particle number projection technique~\cite{Simenel2010PRL,Sekizawa2013PRC}.

Let $V$ denote the region of interest, and its complementary space is $\bar{V}$. The particle number projection operator for neutrons ($q= n$) or protons ($q=p$) in region $V$ reads~\cite{Simenel2010PRL,Sekizawa2013PRC}
\begin{equation}
\hat{P}_m^{(q)}=\frac{1}{2 \pi} \int_0^{2 \pi} \text{d} \theta e^{i\left(m-\hat{N}_V^{(q)}\right) \theta}.
\end{equation}
$\hat{N}_V^{(q)}$ is the particle number operator in the region $V$, defined as
\begin{equation}
\hat{N}_V^{(q)}=\int_V d \boldsymbol{r} \sum_{i=1}^{N^{(q)}} \delta\left(\boldsymbol{r}-\boldsymbol{r}_i\right)=\sum_{i=1}^{N^{(q)}} \Theta_V\left(\boldsymbol{r}_i\right),
\end{equation}
where the Heaviside function divides the space as follows
\begin{equation}
\Theta_V(\boldsymbol{r})=\left\{\begin{array}{lll}
1 & \text { if } & \boldsymbol{r} \in V, \\
0 & \text { if } & \boldsymbol{r} \notin V .
\end{array}\right.
\end{equation}
By applying the particle number projection operator to the total wave function $\Psi(\boldsymbol{r})$, the specific component with $N$ neutrons and $Z$ protons can be extracted:
\begin{equation}
\left|\Psi_{N, Z}\right\rangle=\hat{P}_N^{(n)} \hat{P}_Z^{(p)}|\Psi\rangle .
\end{equation}
Correspondingly, the probability of producing a reaction product consisting of $N$ neutrons and $Z$ protons, is computed as
\begin{equation}
P_{N, Z}=\left\langle\Psi_{N, Z} \mid \Psi_{N, Z}\right\rangle=P_N^{(n)} P_Z^{(p)} .
\end{equation}
Here, $P_N^{(n)}$ and $P_Z^{(p)}$ are the individual probabilities for $N$ neutrons and $Z$ protons, respectively,
\begin{equation}
P_m^{(q)}=\frac{1}{2 \pi} \int_0^{2 \pi} e^{i m \theta} \operatorname{det} \mathcal{B}^{(q)}(\theta) \text{d} \theta,
\end{equation}
where
\begin{equation}
\left(\mathcal{B}^{(q)}(\theta)\right)_{i j}=\langle\psi_i^{(q)} \mid \psi_j^{(q)}(\theta)\rangle,
\end{equation}
and
\begin{equation}
\psi_j^{(q)}(\boldsymbol{r}, \theta)=\left[\Theta_{\bar{V}}(\boldsymbol{r})+e^{-i \theta} \Theta_V(\boldsymbol{r})\right] \psi_j^{(q)}(\boldsymbol{r}) .
\end{equation}

The probability $P_{N, Z}(E, b)$ of forming a fragment with $N$ neutrons and $Z$ protons within the division volume $V$ 
is determined for specific values of the incident energy $E$ and impact parameter $b$. The corresponding cross section for a given channel is subsequently obtained by integrating $P_{N, Z}(E, b)$ over the full range of impact parameters
\begin{equation}
\sigma_{N, Z}(E)=2 \pi \int_{b_{\min }}^{b_{\max }} b P_{N, Z}(E, b) \text{d} b,
\end{equation}
where $b_{\text {min }}$ is the minimum impact parameter for a binary reaction, inside which fusion reactions take place, and $b_{\max }$ is the cutoff impact parameter.

%--------------------------------------------------------------
\subsection{Fragment excitation energy and angular momentum}
%--------------------------------------------------------------

For each reaction channel, the corresponding wave function $|\Psi_{N,Z}\rangle$ is used to compute the total angular momentum $\boldsymbol{J}$ and excitation energy $E^{\ast}$ of the reaction product, which are inputs of the statistical model
calculation~\cite{Sekizawa2014PRC}.

The expectation value of the total angular momentum of a reaction product composed of $N$ neutrons and $Z$ protons is defined as
\begin{equation}
\label{pnp_ave_angular}
\boldsymbol{J}_{N, Z}=\frac{\left\langle\Psi_{N, Z}\right| \hat{\boldsymbol{J}}_V\left|\Psi_{N, Z}\right\rangle}{\left\langle\Psi_{N, Z} \mid \Psi_{N, Z}\right\rangle}=\boldsymbol{J}_N^{(n)}+\boldsymbol{J}_Z^{(p)} .
\end{equation}
This quantity satisfies the identity
$\langle\Psi| \hat{\boldsymbol{J}}_V|\Psi\rangle=\sum_{N, Z} P_{N, Z} \boldsymbol{J}_{N, Z}$, where $\hat{\boldsymbol{J}}_V$ is the total angular momentum operator in region $V$
\begin{equation}
\hat{\boldsymbol{J}}_V=\sum_{i=1}^A \Theta_V\left(\hat{\boldsymbol{r}}_i\right) \hat{\boldsymbol{j}}_i,
\end{equation}
and
\begin{equation}
\hat{\boldsymbol{j}}_i=\left(\hat{\boldsymbol{r}}_i-\boldsymbol{R}_{\text {c.m. }}\right) \times \hat{\boldsymbol{p}}_i+\hat{\boldsymbol{s}}_i.
\end{equation}
$\boldsymbol{R}_{\text {c.m. }}$ is the center-of-mass coordinate of the fragment, $\hat{\boldsymbol{p}}_i$ and $\hat{\boldsymbol{s}}_i$ are the momentum and spin operators, respectively.

The contribution of neutrons $\boldsymbol{J}_N^{(n)}$ ($q=n$) or protons $\boldsymbol{J}_Z^{(p)}$ ($q=p$) is computed from
\begin{equation}
\label{pnp_angular}
\boldsymbol{J}_n^{(q)}=\frac{1}{2 \pi P_n^{(q)}} \int_0^{2 \pi} e^{i n \theta} \operatorname{det} \mathcal{B}^{(q)}(\theta) \sum_{i=1}^{N^{(q)}}\left\langle\psi_i^{(q)}\right| \hat{\boldsymbol{j}}\left|\tilde{\psi}_i^{(q)}(\theta)\right\rangle_V \text{d} \theta,
\end{equation}
where
\begin{equation}
\widetilde{\psi}_i^{(q)}(\boldsymbol{r} , \theta) \equiv \sum_{j=1}^{N^{(q)}} \psi_j^{(q)}(\boldsymbol{r} , \theta)\left(\mathcal{B}^{(q)}(\theta)\right)_{j i}^{-1}.
\end{equation}
The subscript $V$ of the bracket indicates that the spatial integration is performed only over the spatial region $V$.

The energy of a reaction product composed of $N$ neutrons and $Z$ protons is defined as
\begin{equation}
\label{pnp_energy}
E_{N, Z}=\frac{\left\langle\Psi_{N, Z}\right| \hat{H}_V\left|\Psi_{N, Z}\right\rangle}{\left\langle\Psi_{N, Z} \mid \Psi_{N, Z}\right\rangle},
\end{equation}
where $\hat{H}_V$ is a Hamiltonian for the fragment inside the spatial region $V$. There also follows the identity $\langle\Psi| \hat{H}_V|\Psi\rangle= \sum_{N, Z} P_{N, Z} E_{N, Z}$.

In the case of an energy density functional (EDF), $E_{N,Z}$ is given by
\begin{equation}
E_{N,Z} = \frac{1}{4\pi^2 P_N^{(n)} P_Z^{(p)}} \int_0^{2\pi} \int_0^{2\pi} e^{i(N\theta + Z\varphi)} \times \det B(\theta,\varphi) \int_V \mathcal{E}(r,\theta,\varphi)  \text{d}r  \text{d}\theta  \text{d}\varphi,
\end{equation}
where $\det B(\theta, \varphi) = \det B^{(n)}(\theta) \det B^{(p)}(\varphi)$. 
$\mathcal{E}(r, \theta, \varphi)$ denotes the EDF kernel consisting of mixed densities, e.g., $\rho_q(r, \theta) = \sum_{i} \psi_i^{(q)*}(r)\tilde{\psi}_i^{(q)}(r, \theta)$, etc.

The EDF kernel includes the center-of-mass correction~\cite{Bender2000EPJA} in order to subtract the kinetic energy associated with center-of-mass translational motion. Considering $E_{N,Z}$ as the total internal energy of the reaction product, its 
excitation energy can be evaluated as
\begin{equation}
E_{N,Z}^* = E_{N,Z} - E_{N,Z}^{\mathrm{g.s.}},
\end{equation}
where $ E_{N,Z}^{\mathrm{g.s.}}$ is the CDFT ground-state energy of a nucleus with $N$ neutrons and $Z$ protons.

%--------------------------------------------------------------
\subsection{De-excitation process}
%--------------------------------------------------------------

The de-excitation process of primary fragments can be simulated using the statistical model code GEMINI++~\cite{Charity2008}, 
with the initial conditions provided by $ P_{N,Z}, J_{N,Z} $, and $ E^{*}_{N,Z} $. In this model, the excited nucleus undergoes a cascade of binary decays, ceasing when the emission of particles becomes energetically unfeasible or is statistically dominated by $\gamma$-ray emission. GEMINI++ offers a comprehensive treatment of decay channels: light-particle evaporation (including $n, p, d, t$, $^{3}$He, $\alpha$,
$^{6}$He, $^{6-8}$Li, and $^{7-10}$Be) is calculated within the Hauser-Feshbach formulism~\cite{Hauser1952PR}, while the emission of heavier intermediate-mass fragments is described by Moretto's binary decay theory~\cite{Moretto1975NPA}. Additionally, the fission probability is evaluated based on the Bohr-Wheeler formulism~\cite{Bohr1939PR}.

The stochastic nature of the model means that identical initial fragments (with the same $N$, $Z$, $E^{*}$, and $J$) can de-excite through different decay sequences. Therefore, decay probabilities are determined statistically by running a large number of simulations ($N_{\text{trial}}$) for each primary fragment. The probability of the transition $(N,Z) \rightarrow (N',Z')$ is then given by the ratio of the count of outcomes yielding the nucleus ($N'$, $Z'$), denoted $N_{N',Z'}$, to the total number of trials, as defined in
~\cite{Sekizawa2017PRC}: 
\begin{equation}
P_{\text{decay}}(E^{*}_{N,Z}, J_{N,Z}, N,Z; N',Z') = \frac{N_{N',Z'}}{N_{\text{trial}}}.
\end{equation}
The cross section for a secondary reaction product formed after the de-excitation is then evaluated as
\begin{equation}
\tilde{\sigma}_{N',Z'}(E) = 2\pi \int_{b_{\min}}^{b_{\text{max}}} b \, \tilde{P}_{N',Z'}(b,E) \, \mathrm{d}b,
\end{equation}
where $\tilde{P}_{N',Z'}$ denotes the probability that a reaction product composed of $N'$ neutrons and $Z'$ protons is produced after the de-excitation
\begin{equation}
\tilde{P}_{N',Z'} = \sum_{N \geq N'} \sum_{Z \geq Z'} P_{N,Z} P_{\text{decay}}(E^{*}_{N,Z}, J_{N,Z}, N,Z; N',Z').
\end{equation}

%--------------------------------------------------------------

\subsection{Shannon entropy and mutual information}
%--------------------------------------------------------------

The measurement entropy of an observable $\hat O$ can be expressed in terms of the Shannon entropy
\begin{equation}
H[\hat{O}] = -\sum_{x} P(x) \ln P(x),
\end{equation}
where $P({x})$ is the probability distribution of the outcomes $x$ of $\hat O$. In particular, for the observable of the number of nucleons in a fragment after collision, the Shannon entropy of the fragment in the subspace $V$ can be evaluated from
\begin{equation}
    H=-\sum_{N,Z}P_{N,Z} \ln P_{N,Z}. 
    \label{Shannon_entropy}
\end{equation}
The Shannon entropy represents a measure of the uncertainty about $\hat{O}$ before obtaining its value, or 
a measure of the amount of information gained after obtaining $\hat{O}$~\cite{nielsen2010}.

To quantify the correlation between different observables, the mutual information between two observables $A$ and $B$ is defined as~\cite{Auletta2009,CWMa2018PPNP}
\begin{equation}
I(A, B) = H(A) + H(B) - H(A, B),
\end{equation}
where $H(A)$, $H(B)$, and $H(A,B)$ are Shannon entropies computed from the corresponding distributions. If the observables $A$ and $B$ are independent, then the joint probability distribution $P(A,B) = P(A)P(B)$ and the mutual information $I(A,B)$ vanishes. Conversely, two observables are correlated if the mutual information differs from zero. 

%--------------------------------------------------------------

\section{Numerical Details} \label{Sdetail}
%--------------------------------------------------------------

We investigate the $^{40}$Ca + $^{208}$Pb reaction at beam energies $ E_{\text{lab}}$ = 235$-$270 MeV within the framework of TDCDFT. Our calculations use the PC-PK1 relativistic density functional~\cite{Zhao2013PRC}. The initial ground-state configurations of the projectile and target are determined via stationary self-consistent mean-field calculations~\cite{ZXRen2017PRC,DDZhang2022PRC,DDZhang2023IJMPE} on a three-dimensional grid with $ N_x \times N_y \times N_z = 26 \times 26 \times 26$ points and a mesh spacing of 0.8 fm. The collision dynamics are simulated on an enlarged grid of $N_x \times N_y \times N_z = 60 \times 26 \times 60$ points \cite{DDZhang2024PRC}, with the reaction taking place in the x–z plane. The propagation of the single-particle wave functions in time is modeled with a predictor-corrector algorithm, utilizing a fourth-order Taylor expansion for the time-evolution operator and a time step of 0.2 fm/c.

The simulation is initialized with the two nuclei on a Rutherford trajectory, separated by 20$-$24 fm in the center-of-mass frame. The dynamical evolution continues until a termination criterion is satisfied: either the primary fragments separate beyond 20$-$24 fm, or the system fails to separate within 3500 fm/c, indicating a long-lived composite system. The cross section is computed by integrating over a set of impact parameters ($b$). The parameter space is sampled with variable resolution: a coarse set of points (7.5, 8, 9, and 10 fm) is used for $b > 7$ fm, a finer grid with $\Delta b =0.25$ fm for $b < 7$ fm, and an even finer grid of $\Delta b =0.05$ fm near the cutoff $b_{\text{min}}$. For relatively large values of $b$ ($> 7$ fm) the calculated $P_{N,Z}$ do not exhibit significant variations. They change more rapidly in the intermediate region, thus the finer grid, and especially near the cutoff $b_{\text{min}}$.  For impact parameters smaller than $b_{\text{min}}$ fusion occurs. The final cross section is obtained by applying the trapezoidal rule to this dataset.

In the particle number projection (PNP) analysis, a spherical division region $V$ ($R_V =$ 12 fm) is centered on the fragment of interest. The quantities $P_{N,Z}$, $J_{N,Z}$, and $E_{N,Z}$ are computed via numerical integration on a uniform mesh of $M=200$ points. To avoid unphysical energy values that arise for channels with probabilities below $10^{-5}$~\cite{Sekizawa2014PRC,Sekizawa2017PRC}, only products with $P_{N,Z} > 10^{-5}$ are included.

The statistical de-excitation of the reaction products is simulated with the GEMINI++ code \cite{Charity2008}. All calculations use the default model parameters, which are calibrated to reproduce experimental data for a broad spectrum of compound-nucleus masses \cite{Charity2010PRC,Mancusi2010PRC}. For each case, $N_{\text{trial}} = 1000$ Monte Carlo trials were performed.

%--------------------------------------------------------------
\section{Results} \label{Results}
%--------------------------------------------------------------

\begin{figure}[htbp]
	\centering
	\includegraphics[width = 0.85\linewidth]{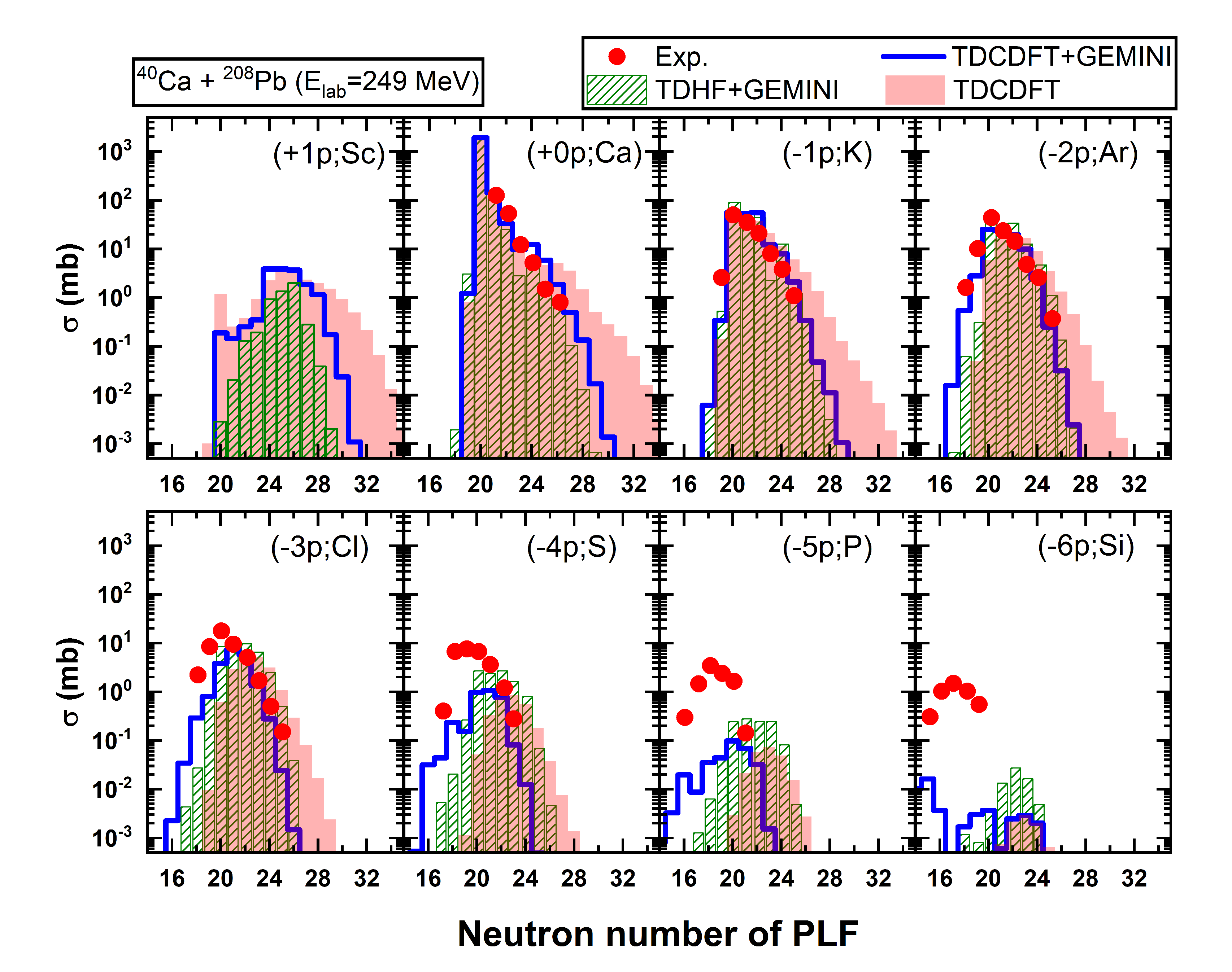}  
  \caption{Cross sections for the projectile-like fragments in the $^{40}$Ca + $^{208}$Pb reaction at $E_{\text{lab}}=249$ MeV. Each panel shows the cross sections for a different proton-transfer channel, as a function of the PLF neutron number (horizontal axis). Dots (red) denote the experimental values \cite{SSzilner2005PRC}. Histograms (pink) present TDCDFT results for primary reaction products, and blue solid lines denote the TDCDFT+GEMINI results for secondary reaction products. For comparison, the TDHF+GEMINI results~\cite{Sekizawa2017PRC} are also shown (green shaded histograms).}
  \label{cross_section}
\end{figure}

In the recent study of Ref.~\cite{DDZhang2024PRC}, the cross sections of primary products from TDCDFT calculations in the $^{40}$Ca + $^{208}$Pb reaction at $E_{\text{lab}}=249$ MeV were analyzed in detail. Using identical model parameters, the maximum impact parameter for fusion was found to be $b = 4.63$ fm, while nucleon transfer becomes negligible for $b \ge 10$ fm. The same range of impact parameters is adopted in the present work.

The cross sections for proton transfer channels to the light fragment in the $^{40}$Ca + $^{208}$Pb reaction at $E_{\text{lab}} = 249$ MeV are shown in Fig.~\ref{cross_section}. The figure's eight panels cover channels from one-proton pick-up ($+1p$) to six-proton stripping ($-6p$), with cross sections displayed as a function of the PLF neutron number. A comparison between the experimental data \cite{SSzilner2005PRC} and the primary products from TDCDFT calculations (filled pink histograms) reveals both agreement and notable discrepancies. While TDCDFT successfully describes the $(+0p)$, $(-1p)$, and $(-2p)$ channels for few-neutron transfers, it overestimates the cross sections for multi-neutron pick-up in these same channels. For more extreme proton-stripping channels ($-3p$ to $-6p$), the theory predicts a much steeper decrease in cross sections and a peak shift to larger neutron numbers compared to the data.

The inclusion of the de-excitation effects through the GEMINI code (TDCDFT+GEMINI, blue solid lines) mitigates these issues. The de-excitation reduces the overestimated cross sections in the ($+0p$) to ($-2p$) channels and corrects the peak positions in the ($-3p$) to ($-6p$) channels, leading to better overall agreement. And we observe a three-peak structure in the ($−6p$) channel, resulting from $\alpha$-particle evaporation from S isotopes and proton evaporation from P isotopes. For comparison, TDHF+GEMINI results \cite{Sekizawa2017PRC} are shown as green histograms. Nevertheless, despite these improvements, the theoretical cross sections for the ($-4p$), ($-5p$), and ($-6p$) channels still deviate from the data by several orders of magnitude. The persistent discrepancy, common to both theoretical frameworks, indicates a fundamental limitation of the mean-field description in quantitatively reproducing these complex multi-nucleon transfer observables.

\begin{figure}[htbp]
  \centering
\begin{minipage}[t]{1\linewidth}  
      \centering
      \includegraphics[width=15cm]{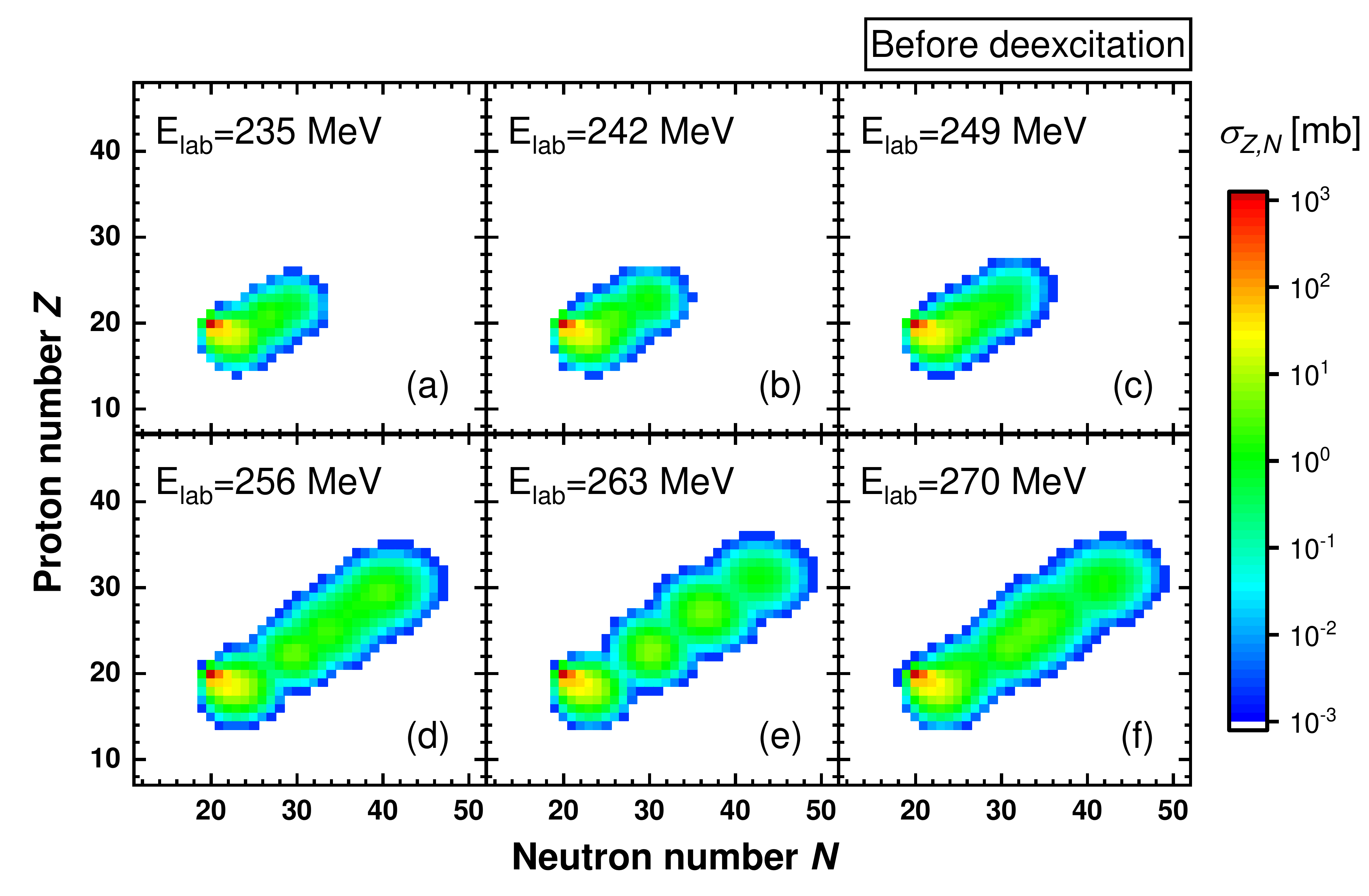}
\end{minipage}
  \begin{minipage}[t]{1\linewidth}
      \centering
      \includegraphics[width=15cm]{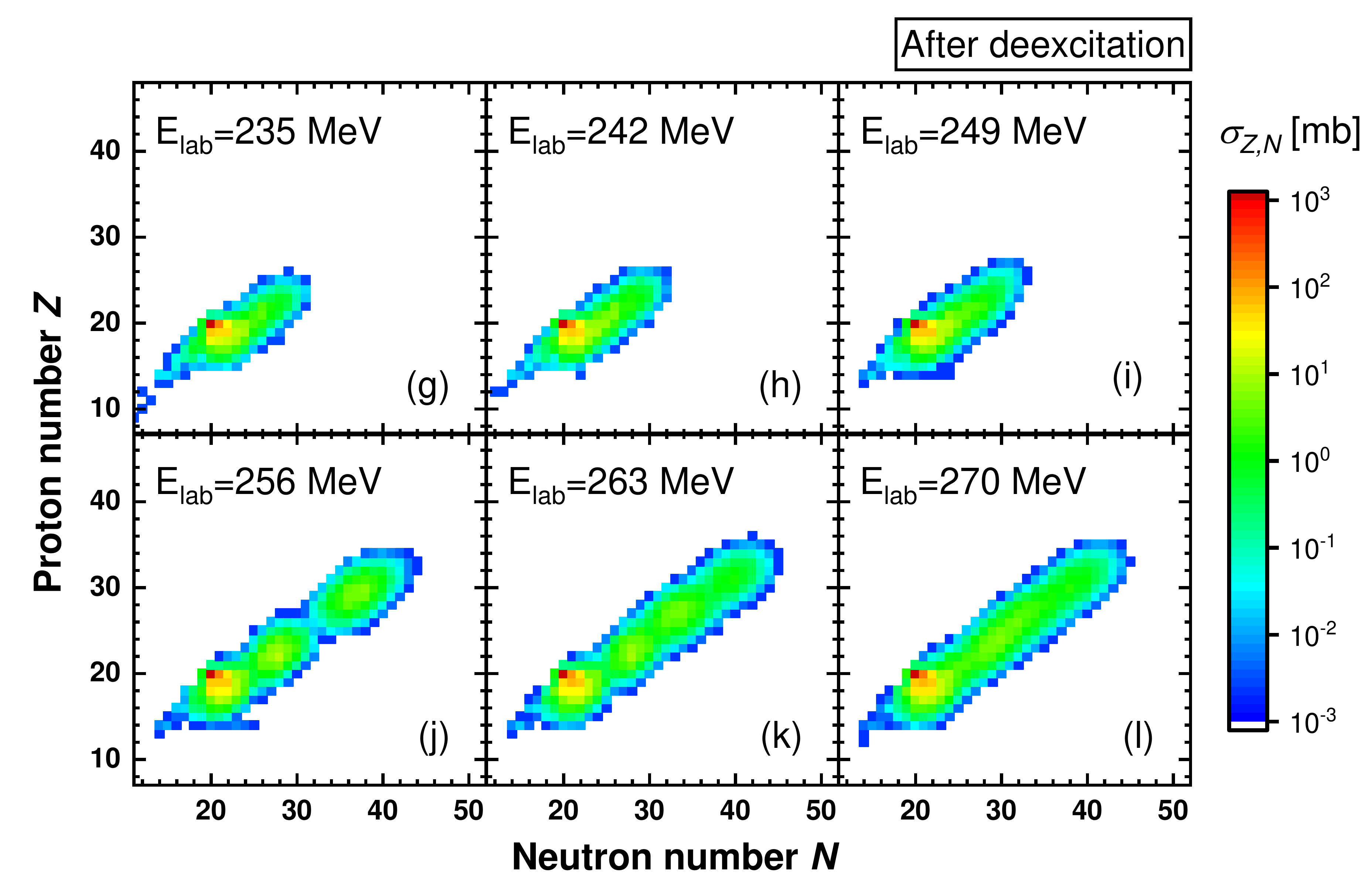}
  \end{minipage}
  \caption{Distribution of cross sections for primary (a)-(f) and secondary (g)-(l) fragments produced in the reaction $^{40}$Ca + $^{208}$Pb, at incident energies between 235 and 270 MeV.}
  \label{dis_cs}
\end{figure}

The cross-section distributions for primary and secondary fragments produced in the $^{40}$Ca + $^{208}$Pb reaction are shown in Fig.~\ref{dis_cs} for a range of incident energies. The calculated maximum fusion impact parameters increase from $b=3.9$ fm to $5.1$ fm as the energy rises from 235 MeV to 270 MeV. The primary fragments exhibit a cross-section distribution peaked at $^{40}$Ca with a pronounced asymmetry; the tail extends more strongly into neutron pick-up channels than into neutron stripping channels. This is a direct consequence of the large initial $N/Z$ asymmetry between $^{40}$Ca ($N/Z=1$) and $^{208}$Pb ($N/Z=1.54$), which drives a charge equilibration process that favors neutron transfer to the projectile. The de-excitation preserves the overall asymmetry but narrows the distribution on the neutron-rich side. This narrowing is attributed to the higher excitation energies typically carried by neutron-rich primary fragments, making them more likely to emit particles and move toward the valley of stability.

As shown in Fig.~\ref{dis_cs} (a)-(f), the cross-section distribution for primary products remains stable as the incident energy increases from 235 to 249 MeV. A pronounced change occurs at 256 MeV, where the distribution expands significantly towards more neutron- and proton-rich products ($N \approx 34 - 44$, $Z \approx 28 - 41$). With further increases to 270 MeV, the distribution stabilizes once more. This pattern persists after the de-excitation, as seen in Fig.~\ref{dis_cs} (g)-(l). Although the decay processes shift the fragments to lower $N$ and $Z$, the abrupt transition between 249 and 256 MeV and the subsequent stability at higher energies remain clearly visible. This indicates that new reaction channels do not open gradually, but rather abruptly once a specific energy threshold is surpassed, remaining active across a broad energy range.

\begin{figure}[htbp]
	\centering
	\includegraphics[width = 0.85\linewidth]{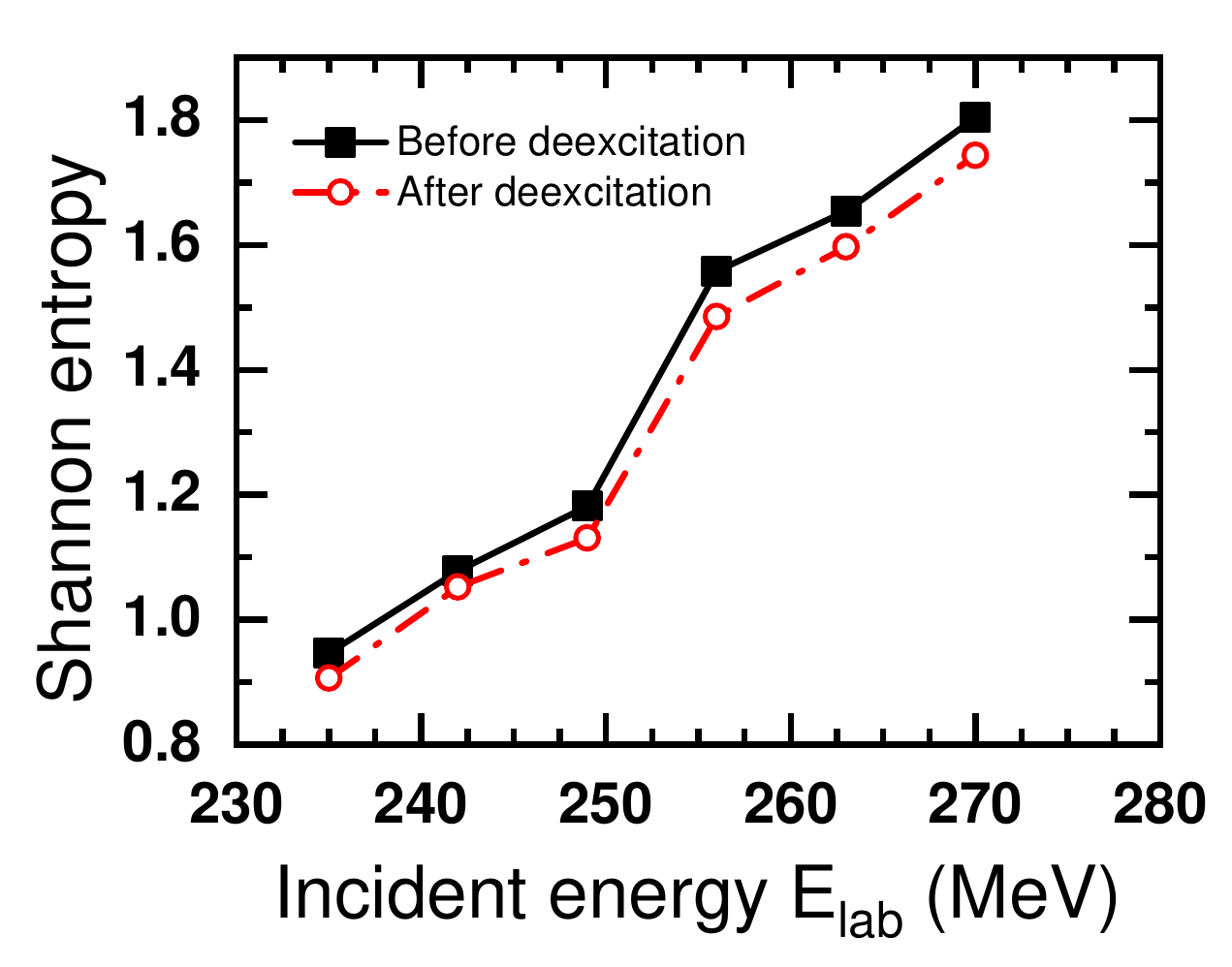}  
  \caption{Cross-section Shannon entropy for primary (black squares) and secondary (red circles) fragments in the reaction $^{40}$Ca + $^{208}$Pb, as a function of the incident energy.}
  \label{entropy}
\end{figure}
%-----------------------------------------------------------------------------------------

We use Shannon entropy to quantitatively describe the distribution of nuclear reaction cross sections, plotting it as a function of incident energy in Fig.~\ref{entropy}. The entropy of secondary fragments is consistently lower than that of primary fragments after the de-excitation. This reduction is expected, as the de-excitation process pulls neutron-rich and proton-rich nuclei back toward the stability line, narrowing the distribution. Furthermore, the Shannon entropy for both primary and secondary fragments increases with incident energy, exhibiting a sharp jump at 249 MeV. This increase stems from two factors: a broader distribution of cross sections within existing channels and the emergence of new reaction channels. The behavior can be broken down as follows: from 235 MeV to 249 MeV, the reaction channels remain largely unchanged, but the cross-section distribution becomes more diffuse, causing a gradual entropy increase. At 256 MeV, a significant number of new channels open, leading to the sharp entropy jump. Beyond this point, with stable channels, the entropy resumes its gradual increase. Thus, a sudden increase in Shannon entropy serves as a clear indicator of new reaction channel openings.

While the analysis has thus far centered on the projectile-like fragments (PLFs) in the $^{40}$Ca + $^{208}$Pb reaction, the target-like fragments (TLFs) has received less attention. This approach is valid for primary fragments, where particle number conservation allows the TLF to be directly determined from the PLF. The subsequent de-excitation process, however, weakens this correlation. Consequently, for secondary fragments, it is essential to investigate the degree to which the PLF-TLF relationship is maintained.

\begin{figure}[htbp]
  \centering
  \begin{minipage}[t]{1\linewidth}  
      \centering
      \includegraphics[width=16cm]{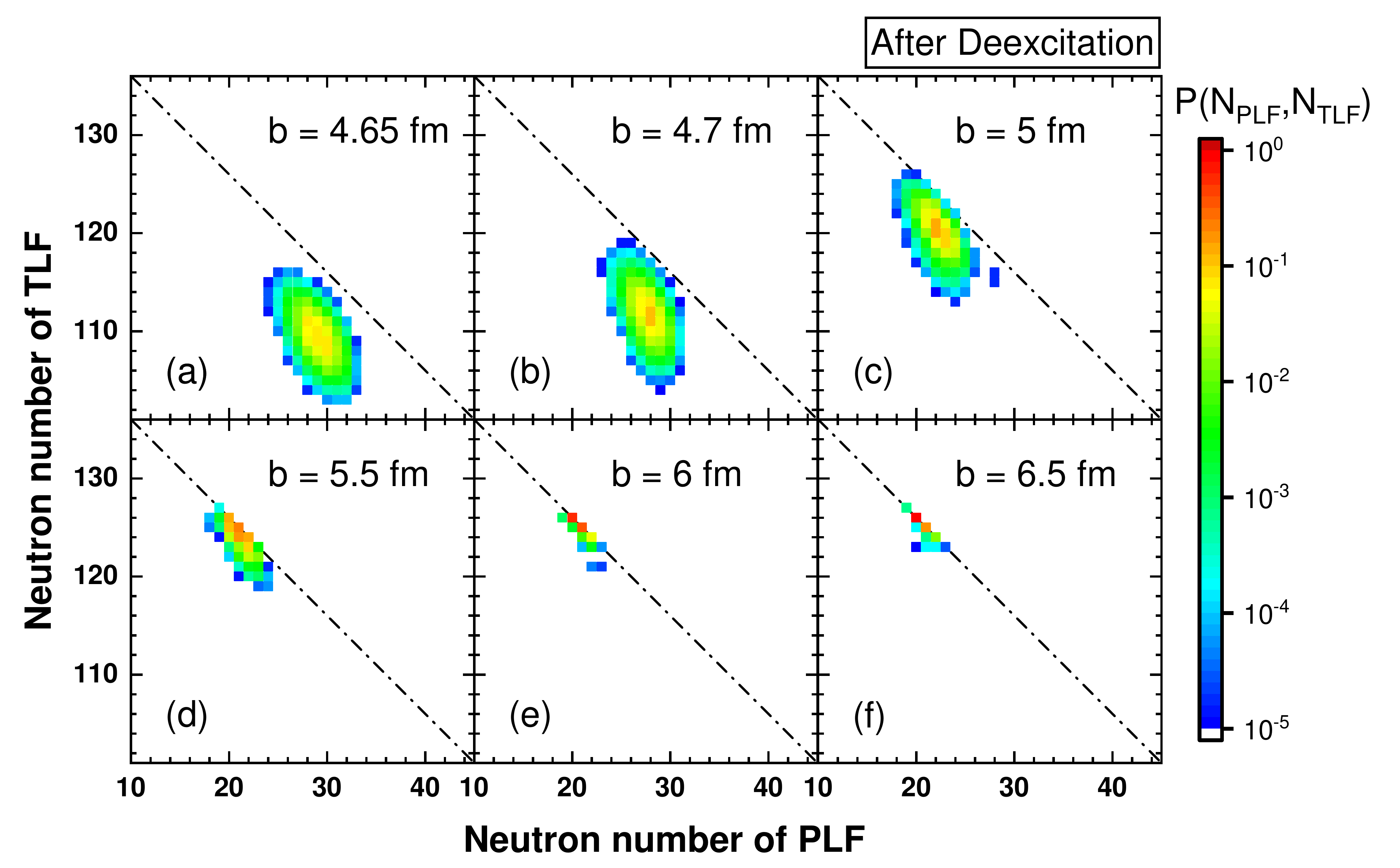}
  \end{minipage}
  \begin{minipage}[t]{1\linewidth}
      \centering
      \includegraphics[width=16cm]{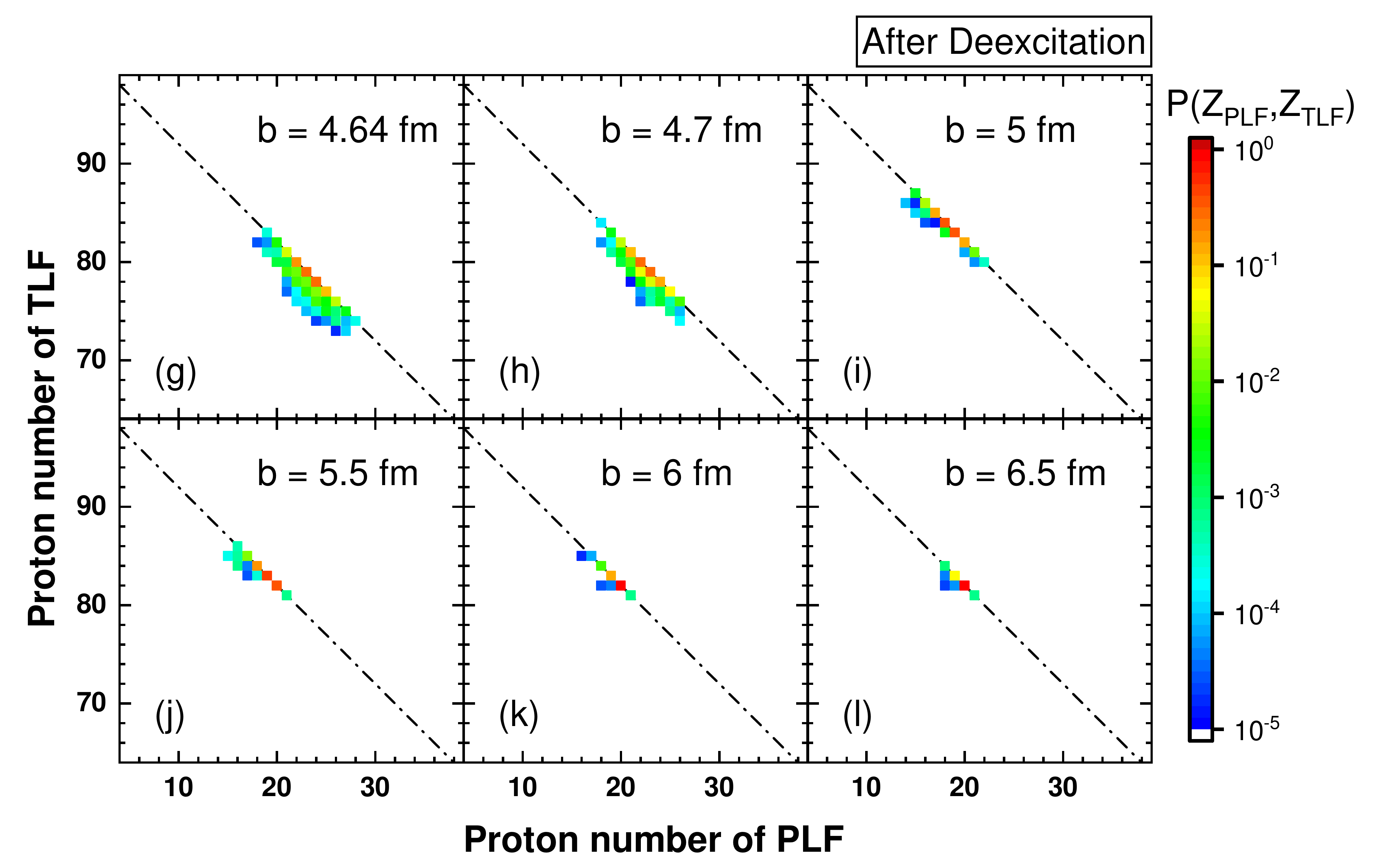}
  \end{minipage}
  \caption{Joint probability distribution of the PLF and TLF following the fragment de-excitation in the $^{40}$Ca + $^{208}$Pb reaction at $E_{\text{lab}}$ = 249 MeV, as functions of the impact parameter, for the neutron number in the upper panels (a)-(f), and the number of protons in the lower panels (g)-(l).}
  \label{joint_pnp}
\end{figure}

To evaluate the PLF-TLF particle number correlation, we computed the joint probability distributions for their neutron and proton numbers. For the post-deexcitation calculation, the TLF decay probability was required; this was obtained using the particle number projection method (Eqs. (\ref{pnp_ave_angular}), (\ref{pnp_energy})) applied to the target-like space $V^{f}_T$, with consistent numerical details, and from simulations via the GEMINI++ code. Before the de-excitation, particle number conservation enforces a strict correlation, constraining the joint probability distribution to a line in the PLF-TLF neutron (proton) number plane. After the de-excitation, this distribution diffuses, as shown in Fig.~\ref{joint_pnp}. The diffusion signifies that the particle number of one fragment can no longer be uniquely determined from the other. This loss of correlation stems from the statistical nature of the de-excitation process: a single primary fragment can decay via multiple pathways into various daughter nuclei, thereby weakening the initial strict correlation. Furthermore, the joint probability distribution for neutron numbers becomes more diffuse with decreasing impact parameter, indicating a greater loss of correlation in more central collisions (Fig.~\ref{joint_pnp} (a)-(f)). A similar diffusion is observed for proton numbers in Fig.~\ref{joint_pnp} (g)-(l). However, the proton distribution exhibits less diffusion than that for neutrons, suggesting that the proton number correlation is better preserved.

\begin{figure}[htbp]
	\centering
	\includegraphics[width = 0.9\linewidth]{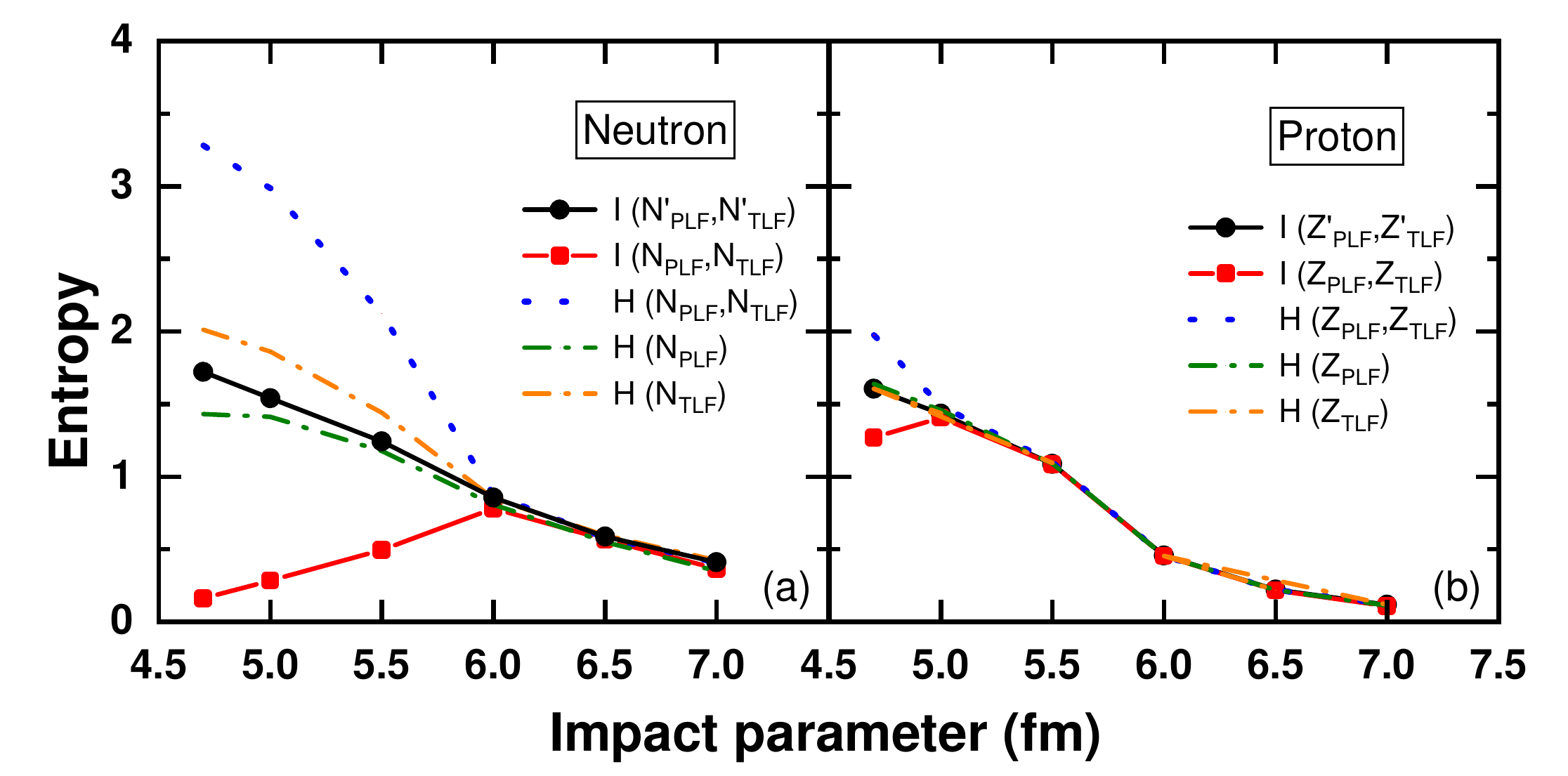}  
  \caption{Mutual information between the PLF and the TLF for (a) the number of neutrons, and (b) number of protons, as functions of the impact parameter. The values of mutual information before and after the de-excitation are denoted by circles (black) and squares (red), respectively, connected by solid lines. The corresponding Shannon entropies after the de-excitation for the PLF (green dot-dashed), the TLF (orange dot-dashed), and their joint probability (blue dashed) are also plotted.}
  \label{mutual_information}
\end{figure}

To quantify the PLF-TLF correlation for proton and neutron numbers, we introduce the mutual information, plotted in Fig.~\ref{mutual_information}. Mutual information measures the information shared between the two fragments, indicating how much knowledge of one fragment reveals about the other. Prior to the de-excitation, particle number conservation dictates a perfect correlation. The mutual information $I(N^{(q)\prime}_{\text{PLF}},N^{(q)\prime}_{\text{TLF}})$ equals the Shannon entropy of either fragment's number distribution $H (N^{(q)\prime}_{\text{PLF}})=H (N^{(q)\prime}_{\text{TLF}})$ where $q$ denotes a neutron ($n$) or proton ($p$). This equality signifies that the information is entirely shared; knowing one fragment completely determines the other. In Fig.~\ref{mutual_information} (a), this pre-deexcitation relationship is represented by circles (black) connected by a solid  for neutron number. The mutual information {\em before} the de-excitation increases with decreasing impact parameter. This trend occurs because closer contact between nuclei leads to more frequent nucleon exchange, resulting in a broader distribution of fragment sizes and thus a larger shared information content.

However, while a stronger interaction increases the initial correlation, it also produces more highly excited fragments. These fragments subsequently undergo more significant de-excitation, which degrades the mutual information. As shown by the squares (red) connected by a solid line in Fig.~\ref{mutual_information} (a), this loss is most pronounced at small impact parameters. Consequently, the {\em post} de-excitation mutual information for neutrons peaks at an intermediate impact parameter of $b=6$ fm. A similar trend is observed for proton number in Fig.~\ref{mutual_information} (b). Crucially, however, the loss of mutual information during the de-excitation is significantly smaller for protons than for neutrons. This strongly indicates that neutron evaporation is the dominant de-excitation mechanism.
%-----------------------------------------------------------------------------------
\section{Summary} \label{Ssum}
%-----------------------------------------------------------------------------------

This study investigates the impact of the nuclear fragment de-excitation on correlation in multi-nucleon transfer (MNT) reactions. It extends our previous works~\cite{BLi2024PRC2,DDZhang2025PLB} , which analyzed entanglement and correlation of spins but were limited to primary fragments -- the highly excited nuclei formed immediately after collision. A critical open question addressed here is how the initial correlation between fragments is altered as these primary fragments decay into the stable secondary products observed experimentally.

To resolve this, we developed a hybrid approach, designated TDCDFT+GEMINI, which integrates two powerful models: time-dependent covariant density functional theory (TDCDFT), to simulate the real-time quantum dynamics of the collision and determine the properties of the primary fragments, and the GEMINI++ statistical model to simulate the subsequent de-excitation cascade of these primary fragments into their final, stable states.

Using the $^{40}$Ca + $^{208}$Pb reaction as a benchmark, our analysis yielded several key findings: (1) Incorporating the de-excitation effects significantly improved the agreement between theoretical cross sections and experimental data, particularly for multi-neutron pickup channels. A notable discrepancy persists for the most extreme transfer channels, underscoring a limitation of the mean-field description. (2) The Shannon entropy of the cross-section distribution, which measures the diversity of reaction products, increases with collision energy. A sharp entropy jump at 256 MeV indicates that new reaction channels open abruptly at a specific energy threshold. (3) We used mutual information to quantify the correlation between the projectile-like (PLF) and the target-like (TLF) fragments. While particle number conservation enforces a perfect correlation before the de-excitation, this correlation is degraded afterward because a single primary fragment can decay into various secondary products. This loss of mutual information is most severe in central collisions (small impact parameters), which produce the most highly excited fragments. The finding that mutual information for proton numbers is better preserved than for neutrons strongly indicates that neutron evaporation is the dominant mechanism responsible for weakening the initial correlation.

In conclusion, this work provides a realistic description of MNT reactions by bridging the initial quantum collision and the final experimental observables. We demonstrate that while the de-excitation process is crucial for reproducing measured cross sections, it also significantly degrades the quantum entanglement and correlations established during the collision, with neutron evaporation playing a pivotal role.

\newpage

\begin{acknowledgments}
This work was supported in part by the High-End Foreign Experts Plan of China,
the National Natural Science Foundation of China (Grants No.12435006, No.12475117, and No.12141501),
the National Key Laboratory of Neutron Science and Technology (Grant No. NST202401016),
the National Key R\&D Program of China 2024YFE0109803,
the National Key Research and Development Program of China 2024YFA1612600,
the High-Performance Computing Platform of Peking University,
the project “Implementation of cutting-edge research and its application as part of the Scientific Center of Excellence for Quantum and Complex Systems, and Representations of Lie Algebras“, Grant No. PK.1.1.10.0004, co-financed by the European Union through the European Regional Development Fund - Competitiveness and Cohesion Programme 2021-2027, 
the Croatian Science Foundation under the project Relativistic Nuclear Many-Body Theory in the Multimessenger Observation Era (IP-2022-10-7773), and the Postdoctoral Fellowship Program and China Postdoctoral Science Foundation under
Grant Number BX20250170.
\end{acknowledgments}

\clearpage
\bigskip
%\bibliography{ref}

\end{document}